\documentclass[manuscript]{aastex631}
\usepackage{mathtools} 
\begin{document}

\title{Properties of Steady Sub-Alfvénic Solar Wind in Comparison with Super-Alfvénic Wind from Measurements of Parker Solar Probe}

\author[0009-0004-4832-0895]{Yiming Jiao}
\affiliation{State Key Laboratory of Space Weather, National Space 
Science Center, Chinese Academy of Sciences, Beijing, China; liuxying@swl.ac.cn}
\affiliation{University of Chinese Academy of Sciences, Beijing, China}
\author[0000-0002-3483-5909]{Ying D. Liu}
\affiliation{State Key Laboratory of Space Weather, National Space 
Science Center, Chinese Academy of Sciences, Beijing, China; liuxying@swl.ac.cn}
\affiliation{University of Chinese Academy of Sciences, Beijing, China}
\correspondingauthor{Ying D. Liu}
\email{liuxying@swl.ac.cn}
\author[0000-0002-8234-6480]{Hao Ran}
\affiliation{State Key Laboratory of Space Weather, National Space 
Science Center, Chinese Academy of Sciences, Beijing, China; liuxying@swl.ac.cn}
\affiliation{University of Chinese Academy of Sciences, Beijing, China}
\author[0009-0005-3941-1514]{Wenshuai Cheng}
\affiliation{State Key Laboratory of Space Weather, National Space 
Science Center, Chinese Academy of Sciences, Beijing, China; liuxying@swl.ac.cn}
\affiliation{University of Chinese Academy of Sciences, Beijing, China}



\begin{abstract}
We identify more than ten steady sub-Alfv\'enic solar wind intervals from the measurements of the Parker Solar Probe (PSP) from encounter 8 to encounter 14.
An analysis of these sub-Alfv\'enic intervals reveals similar properties and similar origins.
In situ measurements show that these intervals feature a decreased radial Alfvén Mach number resulting from a reduced density and a relatively low velocity, and that switchbacks are suppressed in these intervals.
Magnetic source tracing indicates that these sub-Alfv\'enic streams generally originate from the boundaries inside coronal holes, or narrow/small regions of open magnetic fields.
Such properties and origins suggest that these streams are mostly low Mach-number boundary layers (LMBLs), which is a special component of the pristine solar wind proposed by Liu et al. 
We find that the LMBL wind, the fast wind from deep inside coronal holes, and the slow streamer wind constitute three typical components of the young solar wind near the Sun.
In these sub-Alfv\'enic intervals, the Alfv\'en radius varies between 15 and 25 solar radii, in contrast with a typical 12 radii for the Alfv\'en radius of the super-Alfv\'enic wind.
These results give a self-consistent picture interpreting the PSP measurements in the vicinity of the Sun.

\end{abstract}



\section{Introduction} \label{sec:intro}
A sub-Alfv\'enic region is a magnetically dominated region around the Sun where the solar wind speed is slower than the local Alfv\'en speed. 
This region is inside the range of the solar corona and is where the coronal heating and solar wind acceleration occur \citep[e.g.,][]{1980JGR....85.1311H,2019ApJ...877L..35K,2021JPlPh..87c9004C}. 
However, observations of the sub-Alfvénic wind had been limited until the Parker Solar Probe (PSP) mission, which aims to explore the solar wind source region \citep{2016SSRv..204....7F}, sampled the sub-Alfvénic wind for the first time during its 8th solar encounter \citep{2021PhRvL.127y5101K}. 
Studies have been performed on this first sampling as well as a few subsequent intervals of the sub-Alfvénic wind to determine the properties of the coronal plasma \citep[e.g.,][]{2022ApJ...926L..16Z,2022ApJ...926L...1B,2022ApJ...934L..36Z,2022Univ....8..338A,2022ApJ...937...70Z}.
Now with a broader data range available, extensive sub-Alfvénic intervals can be identified from the PSP measurements from encounter 8 to the more recent encounter 14. 
These observations provide the opportunity to analyze the overall properties of the sub-Alfvénic wind in comparison with the super-Alfvénic wind. 
 
As reported by \citet{2021PhRvL.127y5101K}, the first sustained sampling of the sub-Alfvénic wind lasted for 5 hours at a distance of about 20 solar radii from the Sun.
\citet{2021PhRvL.127y5101K} suggest that the first sub-Alfvénic interval is associated with a pseudo-streamer, but the density is usually low.
Further studies showed that the sub-Alfvénic wind detected by PSP is characterized by weaker magnetic field reversals \citep[i.e., switchbacks;][]{2019Natur.576..228K,2019Natur.576..237B} in comparison with the super-Alfvénic solar wind \citep[e.g.,][]{2022ApJ...926L..16Z,2022ApJ...926L...1B,2022ApJ...934L..36Z}.
\citet{2023ApJ...944..116L} interpreted the nature of the sub-Alfvénic intervals as a special type of wind originating from the peripheral areas inside coronal holes, termed as a low Mach-number boundary layer (LMBL).
An LMBL is characterized by an enhanced Alfvén radius, which explains the detection of the sub-Alfvénic wind at a relatively large distance.
\citet{2023ApJ...944..116L} also suggest that switchbacks are naturally suppressed in an LMBL by the low Alfvén Mach number due to their nature as the Alfvénic turbulence.
While this theory is consistent with the observations, it is necessary to examine whether the newly observed sub-Alfvénic solar wind fits this framework.  

The components of the young solar wind beyond LMBLs are also of great interest.  
According to previous studies of the well-evolved solar wind, the solar wind at large distances (e.g., at 1 AU) is divided into fast and slow wind respectively.
The fast wind is relatively homogeneous and is long believed to originate from inside coronal holes \citep[e.g.,][]{1976SoPh...46..303N,1977RvGSP..15..257Z}.
The origins and properties of the slow solar wind, however, are much more diverse.
%some sentences describing the debated sources%
Suggested sources of slow wind include active regions \citep[e.g.,][]{2007Sci...318.1585S,2016SoPh..291..117E,2023A&A...675A..20H}, helmet streamers and pseudo-streamers \citep[e.g.,][]{2000JGR...10525133W,2011ApJ...731..112A,2023ApJ...949...14L}, and coronal hole boundaries \citep[e.g.,][]{1990ApJ...355..726W,2000JGR...10525133W,2005A&A...435..699A,2021MNRAS.508..236W}.
The exact nature of the slow solar wind is still under debate due to various complexities.
For example, some of the slow wind also shows characteristics that are typical of fast wind.
While Alfvénic fluctuations are usually associated with fast solar wind, some slow streams can be highly Alfvénic despite their low velocities \citep[e.g.,][]{1981JGR....86.9199M,2015ApJ...805...84D,2019MNRAS.483.4665D,2020A&A...633A.166P}.
An LMBL also falls into this category, i.e, the Alfvénic slow wind \citep{2023ApJ...944..116L}.
The components of the nascent solar wind near the Sun have not been determined yet.
PSP provides in situ measurements at distances where the solar wind has not yet undergone significant evolution or stream-stream interactions.  
With these measurements, the nascent solar wind can be classified more clearly by their properties and their origins can be inferred.

Another important parameter to be determined is the Alfvén radius $r_A$, which in the classical theory of \citet{1967ApJ...148..217W} is the heliocentric distance where the solar wind become from sub-Alfvénic to super-Alfvénic.
Earlier studies before the PSP mission have estimated $r_A$ to vary from a few to tens of solar radii \citep[e.g.,][]{2019ApJS..241...11C,2018JPlPh..84e7601P,2014ApJ...787..124D,2013ApJ...778..176O,2007ApJS..171..520C}.
Recent works using in situ measurements from PSP have constrained $r_A$ to 10 to 20 solar radii \citep[e.g.,][]{2021ApJ...908L..41L,2023ApJ...944..116L,2023JGRA..12831359B}.
As PSP dives below the Alfvén critical point more times and for longer durations, we have more measurements to constrain the Alfvén radius.  
However, since the plasma properties may change significantly on either side of the Alfvén critical point, previous methods may lead to unrealistically large values for the Alfvén radius for the measurements of the sub-Alfvénic wind.
A new method is needed to give reliable $r_A$ values, especially for the measurements below the Alfvén critical point. 
With the extensive observations from both sides of the Alfvén transition that cover more of the solar longitudes, it is possible to present a more complete and accurate picture of the $r_A$ distribution.

In this paper, we identify steady sub-Alfvénic intervals from PSP measurements during its 8th to 14th solar encounters. 
We analyze the properties of these intervals, trace their solar origins, and categorize them as the same type of solar wind that meets the criteria of an LMBL.
As representatives of the LMBL wind, these sub-Alfvénic streams manifest a different structure in terms of the velocity and density than the commonly known fast wind and the slow streamer wind.
Our result suggests three typical components of the young solar wind where the different streams have not yet mixed.
We also obtain a complete picture of the $r_A$ distribution in the ecliptic plane. 
The contrast of $r_A$ between the current sub-Alfvénic wind and the wind that still remains super-Alfvénic shows the enhancement of $r_A$ in LMBLs.

\section{Analysis and Results} \label{sec:a&r}

\begin{figure}
\includegraphics[scale=0.85]{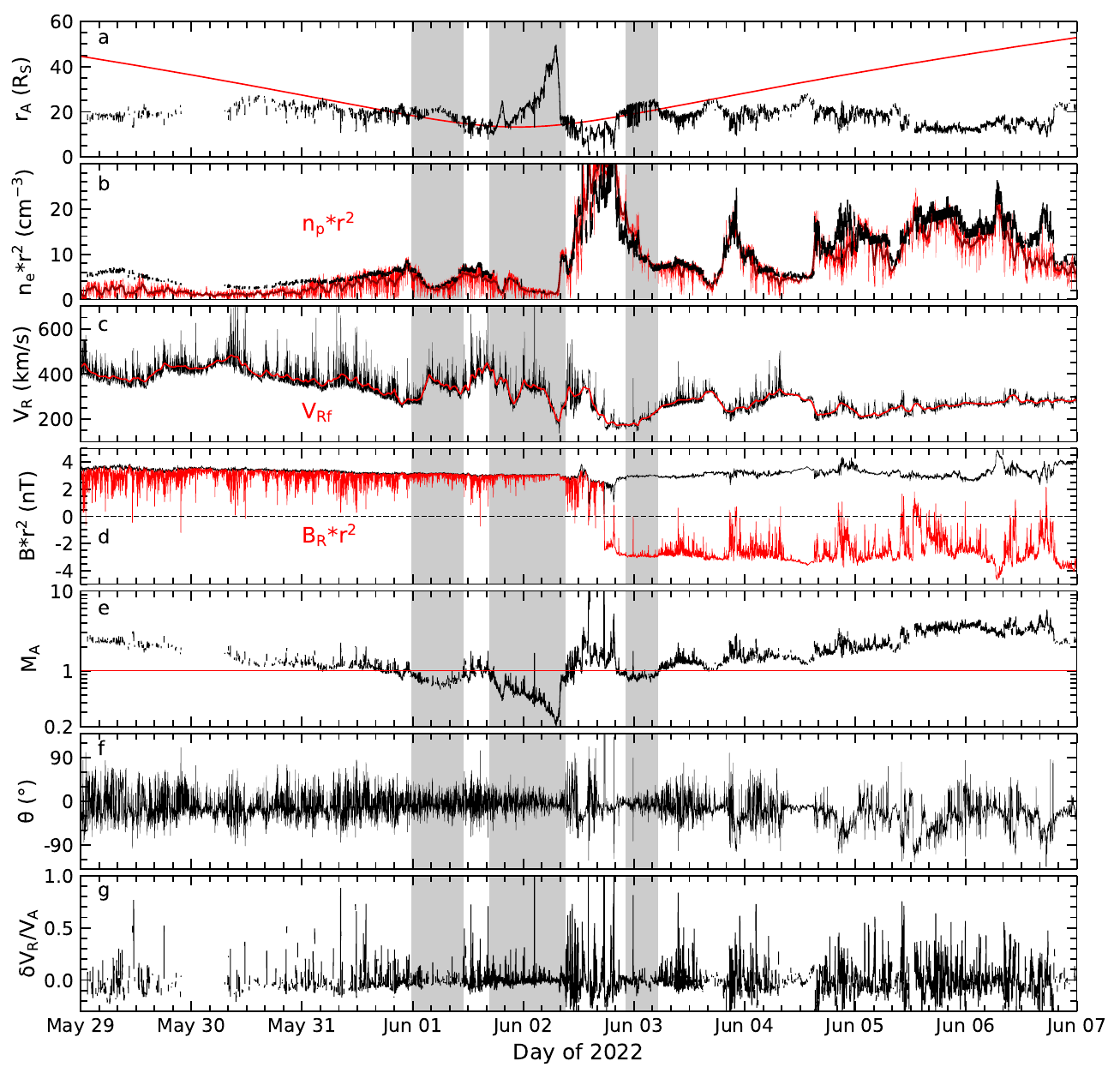}
\caption{PSP measurements at encounter 12. The shaded areas indicate sub-Alfvénic intervals. (a) Alfvén radius compared to the distance of PSP (red). (b) QTN electron density and SPAN-I proton density (red) normalized to 1 AU values, with the filtered proton density values superimposed in dark red. (c) Proton radial velocity and filtered values (red). (d) Normalized magnetic field strength and radial component (red). (e) Radial Alfvén Mach number. (f) Magnetic field deflection angle. (g) Radial velocity enhancement in units of local Alfvén speed.}
\label{fig:1}
\end{figure}

As an example of the measurements, Figure \ref{fig:1} shows the data from encounter 12.
The magnetic field data are obtained from the measurements of the PSP/FIELDS fluxgate magnetometer instrument \citep{2016SSRv..204...49B}. 
The solar wind velocity is measured by the PSP/Solar Probe ANalyzer-Ions (SPAN-I) instruments \citep{2016SSRv..204..131K,2022ApJ...938..138L}. 
The electron density is from quasi-thermal noise (QTN) spectroscopy \citep{2020ApJS..246...44M} and is used as a proxy for the plasma density throughout the study. 
The electron density $n_e$, and magnetic field $B$ are normalized to values at 1 AU by a $1/r^2$ scaling (with r being the heliocentric distance) to eliminate the effect of distance variations. 
All parameters are set to a cadence of 1 minute.

Three sub-Alfvénic intervals are identified at encounter 12 as shown by the shaded areas (see the time periods in Table \ref{tab:1}).
We note a lack of QTN data in the second interval. 
We supplement the data with the proton density from SPAN-I \citep{2022ApJ...938..138L} after applying a low-pass filter to minimize measurement fluctuations.  
The reliability of $n_p$ is verified by comparing it with $n_e$ from QTN (Figure \ref{fig:1}(b)).
A heliospheric current sheet (HCS) crossing was observed between the second and the third intervals.
The sub-Alfvénic intervals mark where the radial Alfvén Mach number $M_A$ keeps lower than 1 for a few hours persistently.
Here $M_A$ is the ratio of the radial solar wind speed to the local Alfvén speed $V_R/V_A$, and $V_A$ is computed as $V_A=B/\sqrt{\mu\rho}$, where $\mu$ is the vacuum magnetic permeability and $\rho$ is the plasma density. 
The Alfvén Mach number compares the kinetic and magnetic energy densities and indicates the entering of the magnetically dominated corona when lower than 1.
PSP observed a transition from a relatively fast, tenuous wind before the sub-Alfvénic intervals to a slow, dense wind after these intervals.
This situation is similar to the cases shown in Figure 1 and Figure 2 of \citet{2023ApJ...944..116L} and suggests that the sub-Alfvénic streams were sampled when the PSP's magnetic footpoint was inside a transition layer.
In these streams, an increased Alfvén speed is caused mainly by a low plasma density since the magnetic field strength shows no significant change.
The solar wind speed $V_R$ also stays low or moderate inside the intervals.
These together lead to the decrease in $M_A$.
Such properties of these sub-Alfvénic intervals satisfy the criteria of an LMBL proposed by \citet{2023ApJ...944..116L}.
Since the LMBL streams have a low Alfvén Mach number compared to other types of wind, their $M_A$ would be the first to decrease below 1 and such wind would be recognized as sub-Alfvénic.
Therefore, the first observed sub-Alfvénic wind is likely to be an LMBL stream.
To further validate this, the solar source of these sub-Alfvénic streams will be traced to determine their origins.

The enhanced magnetic control in these sub-Alfvénic streams also leads to an extended solar corona, or an increased Alfvén radius $r_A$ (Figure \ref{fig:1}(a)).
The Alfvén radius $r_A$ is computed following \citet{2021ApJ...908L..41L} when $M_A$ is close to or exceeds 1. 
This method assumes that the solar wind velocity does not change much between the observational site and the Alfvén critical point. 
Although the approximation is valid in the super-Alfvénic wind (see \citealp{2021ApJ...908L..41L} for details), the solar wind can accelerate a lot well below the Alfvén surface before it reaches the critical point. 
Such a situation may result in a significant error in the above method of $r_A$ calculation when $M_A$ is much less than 1.
In this case, we calculate $r_A$ from the expression 
$L_p+L_m=\Omega r_A^2$
that relates the solar wind angular momentum to $r_A$ \citep{1967ApJ...148..217W,1984JGR....89.5386M,2020ApJS..246...24R,2021ApJ...908L..41L} when $M_A<0.8$.
The threshold of 0.8 is chosen based on our experience although arbitrary.
Because there are complications associated with the measurements of the transverse velocity (see \citealp{2021ApJ...908L..41L} for detailed discussions), we drop the particle term $L_p$ and use the field term only, i.e., $\Omega r_A^2\simeq -(rB_RB_T/\mu \rho V_R)=L_m$, where $B_R$ and $B_T$ are the radial and azimuthal components of the magnetic ﬁeld and $\Omega$ is the solar rotation rate.
Therefore, this can be considered as a lower limit in general, but note that in fast wind the particle term can be negative \citep{2021ApJ...908L..41L,1984JGR....89.5386M}.
Thus, the overall expression for $r_A$ is 
\begin{equation}\label{1}
    r_A\simeq
    \begin{dcases} 
        \frac{r}{M_A},& M_A\geq0.8\\
        (-\frac{rB_RB_T}{\mu\rho V_R \Omega})^{1/2},& M_A<0.8
    \end{dcases}
\end{equation}
This combination of the calculation methods is supposed to give a more reliable $r_A$ for both above and under the Alfvén surface. 
As shown in Figure \ref{fig:1}(a), $r_A$ increased to over 20 radii during the sub-Alfvénic intervals, which indicates an extended corona and facilitates PSP's crossings of the Alfvén surface.

In previous studies, suppressed switchbacks were found in the sub-Alfvénic solar wind \citep[e.g.,][]{2022ApJ...926L...1B,2022ApJ...934L..36Z,2022ApJ...926L..16Z} and in LMBLs \citep{2023ApJ...944..116L}.
A similar phenomenon is also observed in the sub-Alfvénic intervals during encounter 12.
As shown in Figure \ref{fig:1}(d), the magnetic field is almost radial with switchbacks indicated by the spikes in $B_R$, and the occasional changes of signs denote deflections over $90^\circ$. 
However, these large-angle deflections can hardly be seen in the sub-Alfvénic intervals.
To further address the question, we use the parameters $\theta$ and $\delta V_R/V_A$ computed using the methods of \citet{2023ApJ...944..116L}.
The deflection angle $\theta$ is the magnetic field's deviation angle from the radial or anti-radial direction depending on the field polarity, and its sign indicates the deflection direction (Figure \ref{fig:1}(f)). 
Inside the sub-Alfvénic intervals, $\theta$ shows a smaller absolute value than the neighboring super-Alfvénic wind.
We obtain $\delta V_R$ by subtracting a low-pass filtered radial velocity $V_{Rf}$ (Figure \ref{fig:1}(c)) from $V_R$.
The value of $\delta V_R/V_A$ increases with the magnetic field deflection angle by the relation $\delta V_R/V_A=1-\cos\theta$ in Alfvénic fluctuations \citep{2023ApJ...944..116L}. 
It is also reduced in the sub-Alfvénic intervals.
The association of the suppressed switchbacks to a decreased $M_A$ in these sub-Alfvénic LMBLs, as well as in super-Alfvénic LMBLs (see Figure 1 of \citealp{2023ApJ...944..116L} for example), is supportive of the switchbacks' nature as Alfvénic turbulence.

\begin{figure}
\plotone{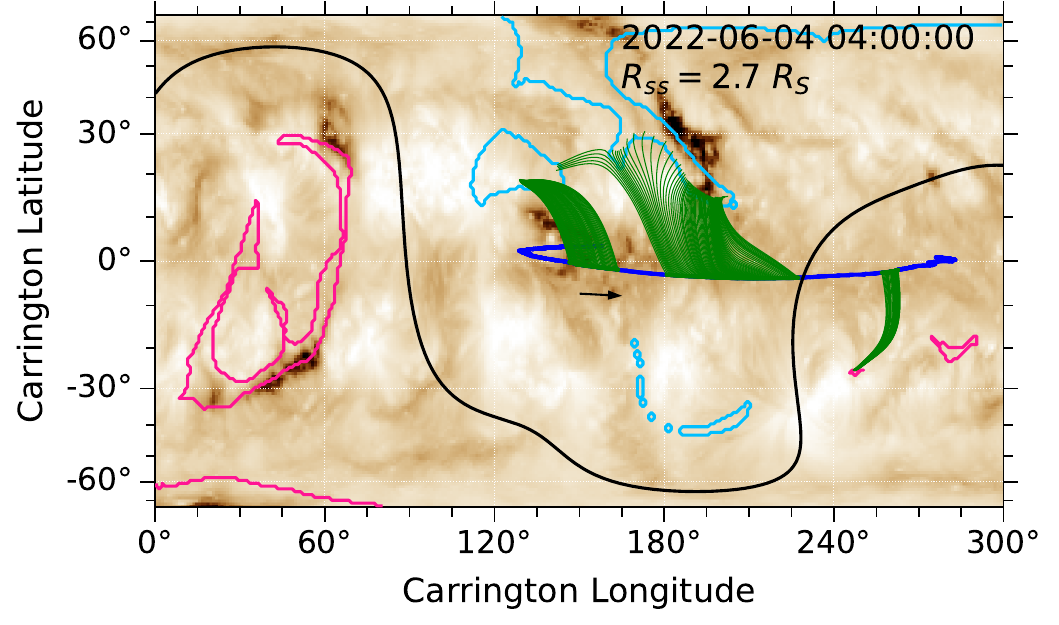}
\caption{Magnetic mapping of the sub-Alfvénic intervals of encounter 12. The connectivity from the PSP projected trajectory to the photospheric sources in the selected intervals is indicated by the three segments of green lines. The light blue and pink contours outline the coronal holes of outward and inward polarities, respectively. The blue curve is the PSP trajectory projected to the source surface, with the direction indicated by the black arrow. The black line indicates the position of the coronal base of the HCS. An SDO EUV synoptic map is used as a background. The time of the ADAPT-GONG magnetogram and the height of the source surface used in the model are also shown.}
\label{fig:2}
\end{figure}

Figure \ref{fig:2} shows the magnetic field source tracing results for the sub-Alfvénic intervals at encounter 12. The magnetic field lines from the spacecraft are ballistically mapped following the Parker spiral field to the source surface at which the magnetic field is set to be radial. Then a potential field source surface (PFSS) model is used to trace the coronal magnetic fields from the source surface to the photospheric sources \citep[e.g.,][]{1969SoPh....9..131A,1969SoPh....6..442S,1992ApJ...392..310W,2020ApJS..246...23B}. 
The construction is based on the Air Force Data Assimilative Photospheric Flux Transport (ADAPT) magnetograms provided by the Global Oscillation Network Group (GONG).
The ADAPT-GONG synoptic map is updated every two hours.
The open field areas given by the PFSS model are compared with EUV imaging observations of coronal holes.
We use Solar Dynamics Observatory (SDO)/Atmospheric Imaging Assembly (AIA) 193 \AA{} synoptic maps of the corresponding Carrington rotations for such a comparison. 
The time of the ADAPT-GONG magnetogram and the height of the source surface are adjusted to best fit PSP and SDO/AIA observations.
The uncertainties of this magnetic mapping were discussed \citep{2020ApJS..246...23B,2021PhRvL.127y5101K,2021A&A...653A..92P}, and the final position error on the photosphere is usually a few degrees according to our experience \citep[e.g.,][]{2022RAA....22c5018M,2021ApJ...921...15C,2023ApJ...944..116L}.

As can be seen from Figure \ref{fig:2}, the magnetic mapping for the first two sub-Alfvénic intervals shows connections ﬁrst to the edge of a low-latitude coronal hole and then to the edge of the equatorial extension of a northern polar coronal hole.
After a crossing of the HCS, the third interval is connected to a small low-latitude coronal hole with a negative polarity.
The crossing of the HCS is consistent with the $B_R$ polarity reversal between the second and third sub-Alfvénic intervals in Figure \ref{fig:1}.
The coronal hole boundaries where the magnetic flux tube expands rapidly have been one of the suggested sources of the slow solar wind \citep[e.g.,][]{2000JGR...10525133W}.
The expansion rate of the magnetic flux tube from the photosphere to the source surface is evaluated by the expansion factor, which is inversely correlated to the wind speed \citep{1990ApJ...355..726W}.
The expansion factor tends to be large in areas just within the coronal hole boundaries, and this would result in a slow solar wind compared to areas deeper inside a coronal hole.
\citet{2023ApJ...944..116L} reasoned that the low wind speed and the tenuous nature of the coronal hole-originated wind together lead to the decreased radial Alfvén Mach number of the LMBL wind coming from coronal hole boundaries.
In addition, narrow strip-shaped coronal holes, such as the one the second interval is connected to, and very small coronal holes, such as the one the third interval is connected to, may also be associated with a large expansion factor of the open field lines.
Such coronal holes could also be the source regions of LMBLs.
The mapping results of the origins of the sub-Alfvénic intervals at encounter 12 satisfy what is expected for an LMBL wind.

We repeat the same procedure of identification and examination of the sub-Alfvénic intervals as well as source tracing for encounters 8 to 14. 
Table \ref{tab:1} lists all the sub-Alfvénic intervals at these encounters.
We select only robust sub-Alfvénic intervals that last over 3 hours. 
Intervals that are too short may be affected by uncertain factors such as temporal solar wind fluctuations and PSP measurement errors.
The longest duration is about 22 hours, which indicates that PSP was steadily below the Alfvén surface. 
Due to the absence of QTN $n_e$ measurements for encounter 11, we replace it with the proton density from SPAN-I.
However, considering the inaccuracy that the replacement may cause, we list the cases of encounter 11 here only for display purpose and remove them in the hereafter statistical analysis. 

The sub-Alfvénic intervals in Table \ref{tab:1} show similarities in terms of their sources and properties.
Source tracing indicates that these intervals are generally connected to the peripheries inside coronal holes, or narrow/small regions of open field lines.
Combined with in situ observations, this result confirms their common nature as LMBLs.
In general, the wind speed of the sub-Alfvénic intervals is low (from 100 to 300 km/s), which can be explained by the fast expansion of magnetic flux tubes from their sources.
The density is also generally small, which is consistent with their origin from within coronal holes.
The magnetic field deflection angle is significantly reduced ($\lesssim 15^\circ$), so instead of being literally ``switchbacks" small deflections are observed in the sub-Alfvénic intervals.
The velocity enhancement in units of the Alfvén speed is also reduced accordingly.
The Alfvén radius is calculated from Equation (\ref{1}) and shows an average value of around 20 $R_S$, which explains the first detection of the sub-Alfvénic wind even at a distance of about 20 $R_S$ from the Sun \citep{2021PhRvL.127y5101K}. 
Note that the Alfvén radius of interval No.11 is abnormally large, so it cannot be considered as a typical case.
More general results are given by other intervals.

The last two rows of Table \ref{tab:1} contrast the assembled sub-Alfvénic wind with the super-Alfvénic wind. 
The analyzed sub-Alfvénic wind data set is a combination of the intervals listed in Table \ref{tab:1} except the intervals from encounter 11 (No.5 and No.6).
The super-Alfvénic wind data is from full observations from encounters 8 to 14 with encounter 11 excluded, and we also exclude transient structures (such as coronal mass ejections). 
Such a data selection should represent a wide range of super-Alfvénic wind properties.
The contrast shows the universal characteristics, such as the low density, low speed, suppressed switchbacks, and enhanced Alfvén radius, for the sub-Alfvénic wind compared with the super-Alfvénic wind.
This distinction also differentiates LMBL streams from the ordinary solar wind.

\begin{figure}
\plotone{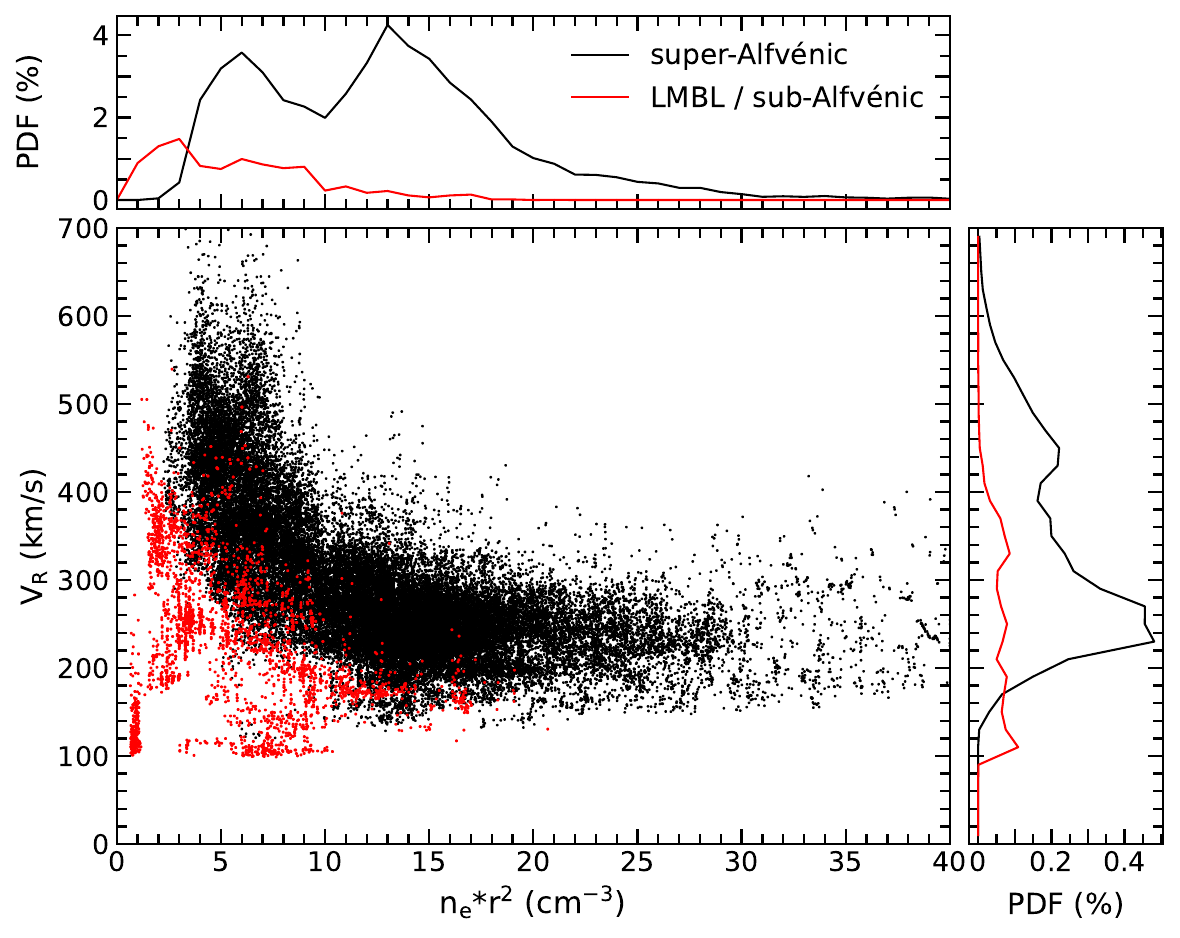}
\caption{Three typical components of the young solar wind as revealed by the distribution as a function of radial velocity and normalized density. The top and right panels give the probability density function (PDF) of the normalized density and velocity. Red is for the LMBL/sub-Alfvénic wind and black is for the super-Alfvénic wind. The PDF of the LMBL/sub-Alfvénic solar wind is amplified by a factor of 3.}
\label{fig:3}
\end{figure}

Using the data selected above, we show in Figure \ref{fig:3} the solar wind distribution as a function of the normalized plasma density and radial velocity for the sub- and super-Alfvénic solar wind.
The correspondence between the solar wind speed and density can be used as a signature to examine the components of the nascent solar wind in the vicinity of the Sun, where the different solar wind streams have not yet interacted considerably.
The solar wind is generally divided according to velocity into fast and slow wind. 
The fast wind is believed to come from inside coronal holes featuring open magnetic field lines and reduced density \citep[e.g.,][]{1976SoPh...46..303N,1977RvGSP..15..257Z}.
Such a type of solar wind can be seen in Figure \ref{fig:3} as a gathering area of the super-Alfvénic wind that is clearly faster than the rest, with velocities concentrating around 450 km/s (the upper PDF peak in the right panel), and with corresponding low densities concentrating around 6 cm$^{-3}$ (the left peak in the top panel).
The slow wind, however, shows more diversity in properties and its origin is still under debate.
\citet{2000JGR...10525133W} suggests that the slow solar wind has two kinds of origins, one from the rapidly diverging open magnetic field rooted at coronal hole boundaries and the other from the closed magnetic loops of helmet streamers.
This difference in origins is suggested to cause the observed variance in multiple parameters of the slow solar wind \citep[e.g.,][]{2015ApJ...805...84D,2020A&A...633A.166P,2021ApJ...910...63G}.
In general, the plasma escaping the closed field lines of streamers tends to be denser, whereas it is more tenuous when from boundaries within coronal holes.
This difference is also seen in Figure \ref{fig:3}. 
The clustered data points of the super-Alfvénic wind corresponding to the right peak in the top panel and the lower peak in the right panel may indicate the dense, slow plasma ejected from streamers.
In contrast, for the sub-Alfvénic wind, the normalized density is substantially decreased while the wind speed also remains low (corresponding to the single, relatively broad peak in the top and right panels).
This is again consistent with the previously discussed LMBL flows with rapidly diverging open magnetic fields from either coronal hole boundaries or small/narrow regions of open fields. 

Based on the analysis above, we conclude that the usual super-Alfvénic wind fits the dichotomy of the fast wind from the coronal hole interiors and the slow wind from streamers (two peaks in the black curves). Some of the slow, dense wind may also come from active regions, although it could be few in the present measurements. The LMBL wind is a third component of the solar wind, a tenuous yet slow wind from coronal hole edges. 
These three components constitute the young solar wind.

\begin{figure}
\plotone{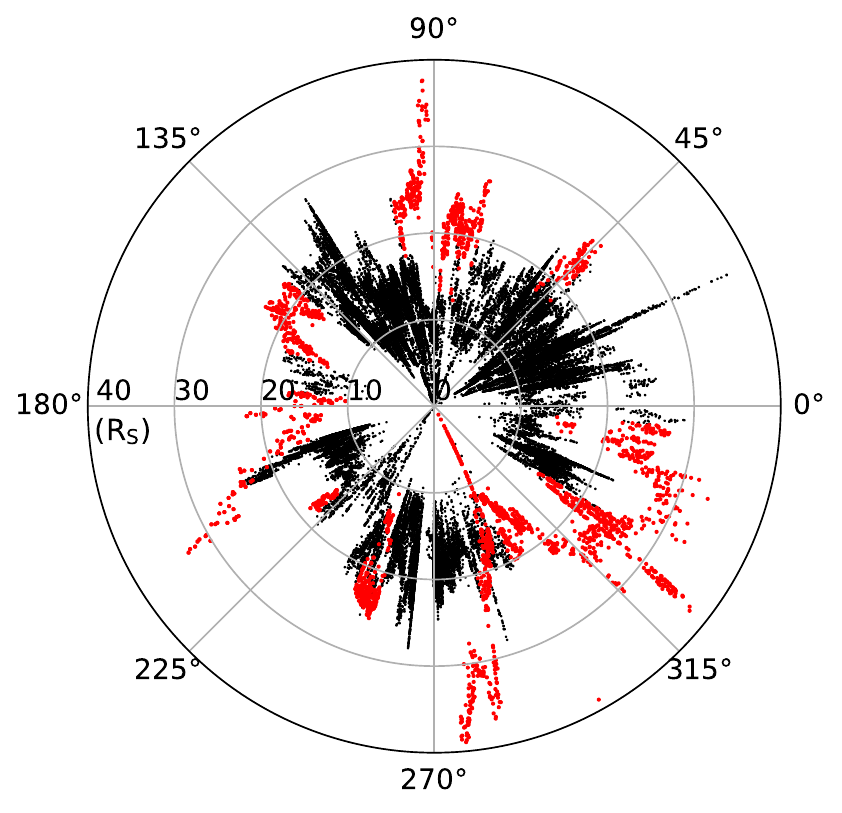}
\caption{Distribution of the Alfvén radius as a function of Carrington longitudes. Red is for the LMBL/sub-Alfvénic wind, and black is for the super-Alfvénic wind. Distances from the center of the Sun are also marked.}
\label{fig:4}
\end{figure}

Figure \ref{fig:4} shows the Alfvén radius distribution as a function of the Carrington longitude of PSP, and the distance from the origin indicates the value $r_A$ at that time.
As the PSP orbit is confined to a few degrees in latitude near the ecliptic, it can be regarded as the configuration of the Alfvén surface in the ecliptic plane.
PSP has circled the Sun from encounters 8 to 14 to give a relatively complete picture.
The height of the Alfvénic radius is shown as varying from 10 to 30 solar radii around the Sun, with the parts associated with the sub-Alfvénic wind generally protruding to larger distances. 
This variation of $r_A$ is consistent with the ``rugged surface" picture \citep{2021ApJ...908L..41L,2023ApJ...944..116L,2023JGRA..12831359B,2021MNRAS.506.4993V}.
PSP's crossings of the protruding parts of the Alfvén surface always correspond to its entry into the LMBL streams.
As a result, PSP sampled solar wind of similar properties and similar origins, as exhibited by the present sub-Alfvénic intervals.
Note that for some longitudes we may see both sub- and super-Alfvénic wind at the same longitudes. 
These are observations at different encounters.

\begin{figure}
\plotone{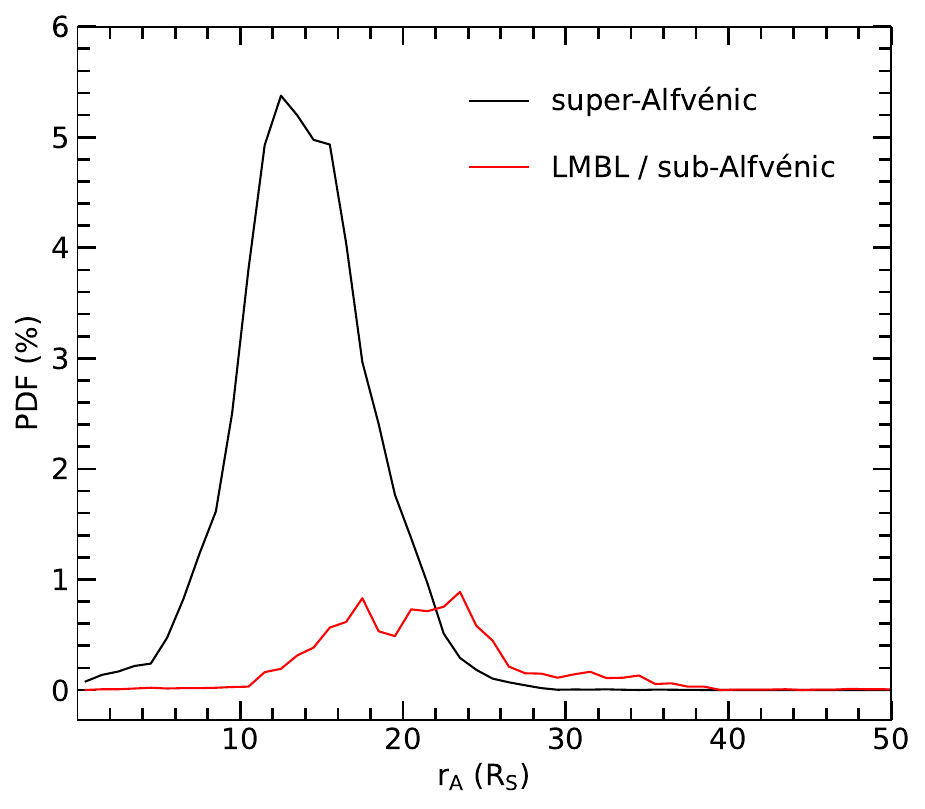}
\caption{Probability density function (PDF) of the Alfvén radius. Again, red is for the LMBL/sub-Alfvénic wind, and black is for the super-Alfvénic wind. The PDF of the LMBL/sub-Alfvénic wind is amplified by a factor of 3.}
\label{fig:5}
\end{figure}

Figure \ref{fig:5} shows the probability density function (PDF) of $r_A$ using data from the 6 encounters.
The PDF corresponding to the super-Alfvénic wind peaks at about 12 solar radii from the center of the Sun.
This value is consistent with the estimate of \citet{2023ApJ...944..116L}.
As for the sub-Alfvénic or LMBL wind, the values of $r_A$ estimated from measurements under the Alfvén critical point are more variant, and the PDF peaks at 15 to 25 solar radii.
These results are also in line with the $r_A$ distribution reported by \citet{2023arXiv231005887C}, which is extrapolated from PSP measurements by assuming the radial trends of the solar wind parameters.
The $r_A$ of the super-Alfvénic wind larger than 15 solar radii corresponds to the protruding parts of the Alfvén surface in Figure \ref{fig:4} that were not crossed by PSP, and the wind remained super-Alfvénic at that moment.
These parts are also likely to be associated with LMBL winds as protrusions of the Alfvén surface are usually simultaneous with drops in $M_A$, which is an important signature of LMBLs.

\section{Conclusions} \label{sec:con}

PSP has observed extensive periods of the sub-Alfvénic solar wind since encounter 8.
We have identified all the steady sub-Alfvénic intervals from the measurements at encounter 8 to encounter 14 and analyzed their origins and properties. 
Together with the super-Alfvénic wind, we have examined key issues including the nature of the sub-Alfvénic wind, the components of the nascent solar wind, and the distribution of the Alfvén radius.
This is the first comprehensive study that includes all the main sub-Alfvénic intervals observed by PSP so far and their comparison with the super-Alfvénic wind.
The main conclusions are summarized as follows.

\begin{enumerate}
\item[1.] 
The observed sub-Alfvénic streams show similarities in their properties and origins, and mostly fall into the category of the special solar wind termed as LMBLs by \citet{2023ApJ...944..116L}.
These sub-Alfvénic streams are characterized by a low or moderate speed and a low density, which result in a decrease in the Alfvén Mach number, an enhancement of the Alfvén radius, and suppression of switchbacks in such winds.
Also, these streams generally originate from the boundaries inside coronal holes or narrow/small regions of open magnetic fields according to the source tracing.
Such properties and origins are consistent with those of an LMBL flow proposed by \citet{2023ApJ...944..116L}. 
LMBLs tend to be the first wind to become sub-Alfvénic as PSP approaches the Sun due to their low Alfvén Mach number.
Until now, observations of the sub-Alfvénic wind are mostly limited to LMBL streams.
As PSP descends to lower perihelia, the spacecraft would sample more frequently the sub-Alfvénic wind, in particular the LMBL wind.
However, note that not necessarily all the sub-Alfvénic wind is an LMBL flow as PSP moves even closer to the Sun.
For example, flows from deeper inside a coronal hole may also have a possibility that their Alfvén surface is crossed given the typical Alfvén radius of about 12 $R_S$ in comparison with PSP's final orbit.
\item[2.] 
The young solar wind is shown to be composed of three typical components, i.e., streamer flows, wind from coronal hole interiors, and wind from coronal hole boundaries (i.e., LMBL flows). 
These three components are revealed by the distribution of the young solar wind with respect to the radial velocity and normalized density. 
The streamer wind is dense and slow as the plasma is trapped by the closed magnetic fields.
In contrast, the wind from coronal hole interiors with open magnetic field lines is tenuous and fast. 
However, the wind speed becomes slower as the magnetic field lines expand faster near the coronal hole boundaries, which gives rise to the LMBL wind (i.e., tenuous and relatively slow).
The streamer wind and LMBL wind together constitute the slow solar wind.
Solar wind originating from active regions could also be dense and slow, although it is not a major part of present measurements and is not discussed in detail in this paper.
\item[3.] 
PSP measurements have provided a complete picture for the Alfvén radius distribution.
We obtain a distribution of the Alfvén radius around the Sun for both sub-Alfvénic and super-Alfvénic wind in the ecliptic plane.
The Alfvén radius of the current super-Alfvénic wind concentrates around 12 solar radii from the center of the Sun.
In contrast, the Alfvén radius of the sub-Alfvénic LMBL intervals is enhanced to 15 to 25 solar radii due to their reduced Alfvén Mach numbers.
These LMBLs are associated with protruding parts of the Alfvénic transition, which facilitates the PSP crossings.
As a result, PSP has observed steady sub-Alfvénic intervals with similar origins and similar properties of the LMBL wind.
\end{enumerate}

%\begin{acknowledgments}
The research was supported by by NSFC under grant 42274201, by the Strategic Priority Research Program of the Chinese Academy of Sciences, Grant No.XDB 0560202, by the National Key R\&D Program of China No.2021YFA0718600, and by the Specialized Research Fund for State Key Laboratories of China. We thank Dr.Huidong Hu for his valuable suggestions. We acknowledge the NASA Parker Solar Probe mission and the SWEAP and FIELDS teams for the use of data. The PFSS extrapolation is performed using the \emph{pfsspy} Python package \citep{2020JOSS....5.2732S}. The data used for PFSS modeling are courtesy of GONG and SDO/AIA.
%\end{acknowledgments}


\bibliography{sample631}{}
\bibliographystyle{aasjournal}


\clearpage

\clearpage

%\clearpage
%\movetabledown=30mm
%\begin{rotatetable}
\begin{deluxetable}{cccccccccccc}
\rotate
\tabletypesize{\small}
\tablewidth{0pt} 
%\tablenum{10}
\tablecaption{Sub-Alfv\'enic Intervals from PSP Measurements at Encounters 8-14.\label{tab:1}}
\tablehead{
\colhead{No.} & \colhead{Enc.}&\colhead{Start} & \colhead{Duration} &
\colhead{$r$} & \colhead{$M_A$}&\colhead{$V_R$} & \colhead{$n_e\cdot r^2$} &
\colhead{$\mid\theta\mid$}&\colhead{$\delta V_R/V_A$}&\colhead{$r_A$} &\colhead{LMBL}\\
\colhead{} & \colhead{}&\colhead{(UT)} & \colhead{(hr)} &
\colhead{$(R_S)$} & \colhead{}&\colhead{(km\,s$^{-1})$} & \colhead{(cm$^{-3})$} &
\colhead{$(^\circ)$}& \colhead{}&\colhead{$(R_S)$} &\colhead{}
} 
\colnumbers
\startdata 
1&8&2021-04-28 09:33&5.2&19.1&0.81$\pm$0.07&318$\pm $47&2.7$\pm $1.2&15$\pm $8&0.04$\pm$0.03& 22.9$\pm $1.9& Y\\
2&9&2021-08-09 21:26&3.0&16.1&0.57$\pm $0.09&159$\pm $21&6.8$\pm $1.5&12$\pm $5&0.03$\pm$0.03&20.8$\pm $2.4&Y\\
3&10&2021-11-21 21:17&3.7&15.6&0.42$\pm $0.10&110$\pm $16&7.6$\pm $1.8&8$\pm $4&0.07$\pm$0.09&20.5$\pm $2.3&Y\\
4&10&2021-11-22 02:38&8.0&18.1&0.58$\pm $0.19&131$\pm $17&8.0$\pm $2.6&14$\pm $8&0.03$\pm$0.03&25.1$\pm $3.9&Y\\
5&11$^*$&2022-02-25 12:46&3.3&13.3&0.76$\pm $0.12$^*$&319$\pm $60&4.8$\pm $1.4$^*$&12$\pm$9&0.03$\pm$0.03$^*$&15.3$\pm $0.8$^*$&Y\\
6&11$^*$&2022-02-25 18:35&4.9&13.7&0.68$\pm $0.16$^*$&315$\pm $42&3.8$\pm $1.0$^*$&13$\pm $8&0.04$\pm$0.04$^*$&17.4$\pm $6.1$^*$&Y\\
7&12&2022-05-31 23:46&11.2&16.5&0.76$\pm $0.09&338$\pm $41&4.4$\pm $1.4&15$\pm $10&0.05$\pm$0.06&18.8$\pm $2.1&Y\\
8&12&2022-06-01 16:38&16.6&13.6&0.52$\pm $0.17&318$\pm $73&5.2$\pm $2.9&12$\pm $8&0.04$\pm$0.08&18.6$\pm $6.4&Y\\
9&12&2022-06-02 22:14&7.0&19.8&0.85$\pm $0.08&198$\pm $26&10.0$\pm $2.4&11$\pm $9&0.03$\pm$0.08&21.3$\pm $3.5&Y\\
10&13&2022-09-06 08:40&8.7&13.9&0.46$\pm $0.14&256$\pm $41&4.1$\pm $1.2&10$\pm $7&0.04$\pm$0.08&22.0$\pm $9.3&Y\\
11\tablenotemark{a}&13&2022-09-06 17:40&19.3&17.8&0.35$\pm $0.26&168$\pm $57&2.4$\pm $1.9&10$\pm $8&0.03$\pm$0.03&52.0$\pm $27.0&Y\\
12&14&2022-12-10 20:02&22.0&14.2&0.58$\pm $0.14&284$\pm $89&7.6$\pm $4.9&14$\pm $9&0.05$\pm$0.07&22.8$\pm $3.8&Y\\
\hline
sub-Alfvénic&{}&{}&{}&{}&{}&246$\pm $95&5.8$\pm $3.8&12$\pm $9&0.04$\pm$0.06&21.0$\pm $5.7\tablenotemark{a}&{}\\
super-Alfvénic&{}&{}&{}&{}&{}&333$\pm $111&13.1$\pm $6.6&33$\pm $23&0.16$\pm$0.29&13.9$\pm $4.1&{}\\
\enddata
\tablecomments{Columns (1-5) correspond to the number, encounter number, start time, duration, and average PSP distance of the intervals, respectively. Columns (6-11) give the mean value and standard deviation of the Alfvén Mach number, radial speed, normalized density, absolute deflection angle, radial speed enhancement in units of the Alfvén speed, and Alfvén radius, respectively. Column (12) indicates whether the interval is an LMBL wind or not (Y/N). The * superscript marks where $n_e$ is replaced by the proton density from SPAN-I as the QTN density is not available.}
\tablenotetext{a}{Here $r_A$ of interval No.11 is discarded for it is an outlier and causes a large deviation in statistical characteristics.}
\end{deluxetable}
%\end{rotatetable}

\clearpage

\clearpage

\clearpage
\end{document}